\documentclass[manuscript]{aastex}

\shorttitle{On the continuum radio-spectrum of Cas A}
\shortauthors{Oni\'c \& Uro\v sevi\'c}

\begin{document}

\title{On the continuum radio-spectrum of Cas A: possible evidence of the non-linear particle acceleration}

\author{D. Oni\'c\altaffilmark{1}}

\email{donic@matf.bg.ac.rs}

\and

\author{D. Uro\v sevi\'c\altaffilmark{1, 2}}

\altaffiltext{1}{Department of Astronomy, Faculty of Mathematics, University of Belgrade, Serbia}
\altaffiltext{2}{Isaac Newton Institute of Chile, Yugoslavia Branch}

\begin{abstract}
Integrated radio-spectrum of Cas A in continuum was analyzed with special emphasis on possible high frequency spectral 
curvature. We conclude that the most probable scenario is that {\it Planck's} new data reveal the imprint of non-linear 
particle acceleration in the case of this young Galactic supernova remnant (SNR).
\end{abstract}

\keywords{ISM: individual (Cas A) -- radiation mechanisms: non-thermal -- acceleration of particles -- 
ISM: supernova remnants -- radio continuum: ISM}

\section{Introduction}

{\it Planck}\footnote{{\it Planck} is a project of the European Space Agency (ESA) with instruments provided by two 
scientific consortia funded by ESA member states, with contributions from NASA (USA) and telescope reflectors provided 
by a collaboration between ESA and a scientific consortium led and funded by Denmark.} is the third generation space 
mission to measure the anisotropy of the cosmic microwave background. It observed the sky in nine frequency bands 
covering 30 - 857 GHz with high sensitivity and angular resolution from $5'$ to $31'$ (Arnaud et al.\@ 2014 and references 
therein). The Low Frequency Instrument covers the 30, 44, and 70 GHz bands while the High Frequency Instrument covers 
the 100, 143, 217, 353, 545, and 857 GHz bands. {\it Planck's} sensitivity, angular resolution, and frequency coverage 
make it a powerful instrument for Galactic and extragalactic astrophysics as well as cosmology.

Cassiopeia A (Cas A) is a very bright and young supernova remnant (SNR), likely due to a historical supernova around
$353-343$ yr ago (Ashworth 1980; Fesen et al.\@ 2006). Its estimated distance is $3.33\pm0.10$ kpc 
(Alarie et al.\@ 2014). Recently, this SNR was observed by {\it Planck} and the results were published 
in Arnaud et al.\@ (2014). It was shown that Cas A is a distinct compact source from 30 - 353 GHz but 
becomes confused with unrelated Galactic clouds at the highest {\it Planck} frequencies (545 and 857 GHz). 
The apparent excess radiation at 217 and 353 GHz is proposed to be due to coincidental peak in the unrelated 
foreground emission or to cool dust in the supernova remnant (Arnaud et al.\@ 2014).

The slightly concave-up forms of radio-spectra were detected for some young SNRs (Reynolds \& Ellison 1992). The main 
reason for high frequency concave-up curvature in radio-spectra of young SNRs should originate in non-linear diffuse 
shock particle acceleration (see Uro\v sevi\'c 2014 and references therein). Due to positive identification of infrared 
synchrotron radiation from Cas A, Jones et al.\@ (2003) gave indication that the radio-spectrum of this SNR should be 
concave-up.

This paper focuses on the analysis of origin of the high-frequency curvature in radio-spectrum of Cas A.

\section{Analysis and Results}

There are various observations of SNR Cas A (Green 2014). Flux densities at different frequencies for SNR Cas A 
were taken from different papers. Different samples for data analysis were used to account for differences 
in data gathered from various literature. The first sample includes data from Arnaud et al.\@ (2014), 
Baars et al.\@ (1977) and references therein as well as from Mason et al.\@ (1999). The original data 
(data for original epochs) are taken from Table 2 of Baars et al.\@ (1977), not the scaled ones. The second 
sample is formed by addition of data from Hurley-Walker et al.\@ (2009) and Liszt \& Lucas (1999). The frequency 
range for both samples is 0.55 - 353 GHz. The integrated flux density of $52\pm7$ Jy at 353 GHz was adopted 
in accordance with the discussion presented in Arnaud et al.\@ (2014).

Before actual analysis, the flux densities were appropriately scaled to account for secular fading. The epoch ($2010.0$) 
is chosen to coincide with {\it Planck's} intermediate astrophysics results (Arnaud et al.\@ 2014). Two scaling relations 
were used in this paper. The first one, commonly used, is taken from Baars et al.\@ (1977) in the following form
\begin{eqnarray}
d(\nu)= a - b \log\nu_{[{\rm GHz}]},\nonumber \\
\Delta d(\nu) = \Delta a + \Delta b\ |\log\nu_{[{\rm GHz}]}|,\nonumber \\
a = (0.0097\pm0.0004),\ b = (0.003\pm0.0004),
\end{eqnarray} where $d(\nu)$ is the secular decrease in the flux density at given frequency and $\Delta d(\nu)$ 
is an appropriate error estimate.

On the other hand, it was debated that this scaling relation is not appropriate (O'Sullivan \& Green 1999) especially 
for the lowest radio-frequencies (Helmboldt \& Kassim 2009 and references therein).

Due to the lack of accuracy of the above mentioned relation, the scaling proposed in Vinyaik\u{\i}n (2014) was also used in 
the following form \begin{eqnarray}
d(\nu)= a - b \ln\nu_{[{\rm GHz}]} - c\ \nu_{[{\rm GHz}]}^{-2.1},\nonumber \\
\Delta d(\nu) = \Delta a + \Delta b\ |\ln\nu_{[{\rm GHz}]}| + \Delta c\ \nu_{[{\rm GHz}]}^{-2.1},\nonumber \\
a = (0.0063\pm0.0002),\ b = (0.0004\pm0.0001),\nonumber \\
c = (0.0151\pm0.0016)\times10^{-5}.
\end{eqnarray}

The results obtained analyzing data formed by these two scaling relation are compared.

The flux density for the desired epoch and the appropriate error estimate are calculated by
\begin{eqnarray}
S_{t_{2}}(\nu)=S_{t_{1}}(\nu)\left(1 - d(\nu)\right)^{T},\ T = t_{2} - t_{1}, \nonumber \\
\Delta S_{t_{2}}(\nu) = S_{t_{1}}(\nu) \left(\frac{\Delta S_{t_{1}}(\nu)}{S_{t_{1}}(\nu)} + T\ \frac{\Delta d(\nu)}{1 - d(\nu)}\right).
\end{eqnarray}

It is worth noting that radio-spectrum of Cas A shows low-frequency cut-off due to thermal absorption (Vinyaik\u{\i}n 2014) 
or possibly synchrotron self-absorption. As the main interest of this paper involves analysis of high frequency part of the 
Cas A radio-spectrum, first, only data at frequencies higher than around 550 MHz were analyzed (for both scaling relation).

\begin{figure*}
\centering
\includegraphics[angle=-90,scale=.50]{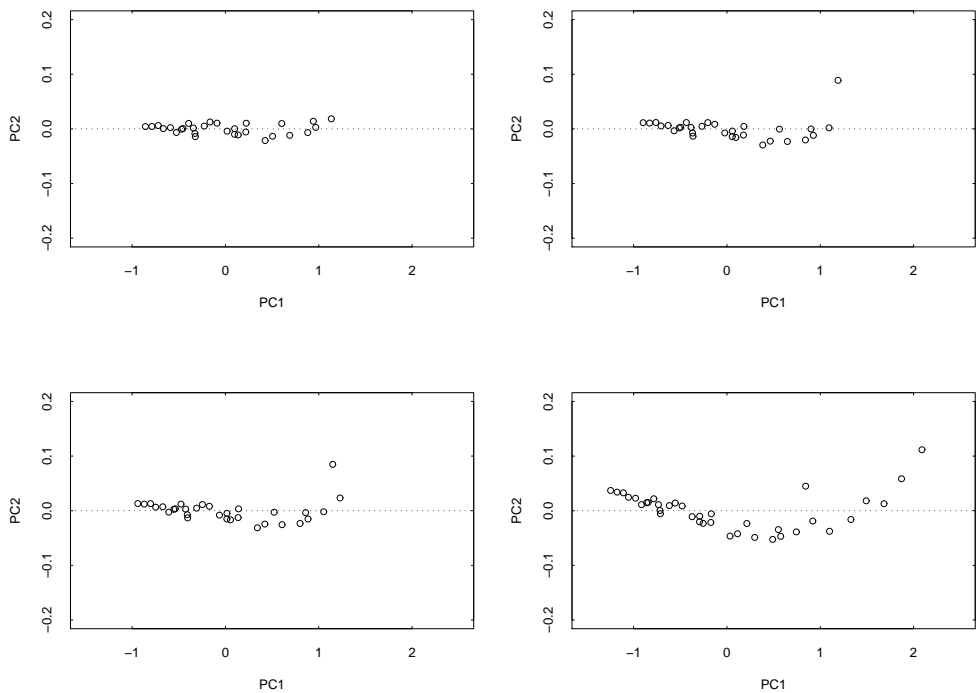}
\caption{The results of PCA on data from the first sample. Scaling relation taken from Baars et al.\@ (1977) was used. 
PC1 and PC2 correspond to the first and second principal component, respectively. In the upper left figure are the results of PCA 
on the radio-data between 550 MHz and 30 GHz. Departure from the linear relationship is obvious when the observation at 
30 GHz is added (upper right figure), both data at 30 and 32 GHz are added (down left figure) and when the whole range 
of frequencies up to 353 GHz is used (down right figure)}\end{figure*}

To show that the observed high-frequency radio-spectrum of Cas A is actually curved, the principal component analysis 
(PCA) was first applied (Babu \& Feigelson 1996). The principal components are the eigenvectors of the covariance 
matrix. Original data are represented in the basis formed by these vectors. The first principal component accounts 
for as much of the variability in the data as possible, i.e.\@ it has the largest possible variance. The second 
principal component has the highest variance possible under the constraint that it is orthogonal to the first one. 
In that sense, PCA includes calculation of the highest data variability direction which can be used to detect 
departures from the pure power-law spectra (linear fit in log-log scale).

In Figure 1, the results of PCA on logarithmically transformed data ($\log\nu, \log S_{\nu}$) from the first 
sample are presented. PC1 and PC2 correspond to the first and second principal component, respectively. The 
plotted values represent the original data in the basis of principal components (zero centered data multiplied by 
the rotation matrix whose columns contain the eigenvectors). The covariance of these values represents a diagonal 
matrix with squares of standard deviations of the principal components eigenvalues as its elements. In the upper 
left figure are the results of PCA on the radio-data between 550 MHz and 30 GHz. It is clear that the radio-spectrum 
follows pure power-law relation. Addition of the observation at 30 GHz (upper right figure), both data at 30 and 32 
GHz (down left figure) and if whole range of frequencies up to 353 GHz is analyzed (down right figure), shows that 
the radio-spectrum of Cas A is positively curved (concave up) above around 30 GHz. The similar results are obtained 
for the radio-data scaled by the relation of Vinyaik\u{\i}n (2014). If data taken from the second sample are used, 
similar results are obtained although the scatter is more prominent. The same conclusions hold if both scaling 
relations are used.

As the high-frequency radio-spectrum of Cas A is clearly curved, it is not appropriate to be fitted by the pure power-law. 
In the most simple case, it can be fitted by the two power-laws, below and above 30 GHz (excluding, of course, the 
low-frequency data influenced by the low-frequency cut-off). On the other hand, it is known that one of the 
repercussions of non-linear effects of particle acceleration is the curved particle spectrum which in turn gives rise 
to the curved synchrotron spectrum (Ellison \& Eichler 1984). In that sense, it is more appropriate to represent 
radio-spectrum of young SNRs with approximative relation that represents varying power-law of the form \begin{equation}
S_{[{\rm Jy}]}(\nu) = S_{[{\rm Jy}]}(1\ \!\!\mathrm{GHz})\ \nu_{[{\rm GHz}]}^{-\alpha + a\log\nu_{[{\rm GHz}]}},
\end{equation}
where $\alpha$ is the standard synchrotron spectral index and $a$ is the parameter of spectral curvature which should be 
positive due to the non-linear behavior of DSA (Houck \& Allen 2006; Vinyaik\u{\i}n 2014 and references therein). Of course, 
more general relation applied to the larger range of energies would include appropriate high-frequency synchrotron spectral 
cut-off which we do not take into account in this analysis.

\begin{table*}[!t]\centering
\caption{Varying power-law best fitting parameters (0.55 - 353 GHz).}
\begin{tabular}{@{}ccccc}
\tableline\tableline
sample (scaling)&$\alpha$&$a$&$\chi^{2}\ (k)$&$R^{2}$\\
\tableline
1 (Baars at al.\@ 1977)&$0.790\pm0.016$&$0.056\pm0.008$&$23.11\ (34)$&$0.997$\\
2 (Baars at al.\@ 1977)&$0.806\pm0.016$&$0.062\pm0.008$&$38.55\ (42)$&$0.995$\\
\hline
1 (Vinyaik\u{\i}n 2014)&$0.841\pm0.014$&$0.065\pm0.007$&$27.33\ (34)$&$0.997$\\
2 (Vinyaik\u{\i}n 2014)&$0.854\pm0.013$&$0.069\pm0.007$&$50.54\ (42)$&$0.995$\\
\tableline
\end{tabular}
\end{table*}

The results of weighted least-square fits to data for this model applied to both samples and for different scaling 
relations are summarized in Table 1. Fit properties are given in 
form of $\chi^{2}$ and $R^{2}$ defined as \begin{eqnarray}
\chi^{2} = \sum_{i=1}^{N}w_{i}\left(y_{i}-f(x_{i}; \beta_{1},...,\beta_{p})\right)^{2},\ w_{i}=\frac{1}{\sigma_{i}^{2}}, \nonumber \\ 
R^{2} = 1 - \frac{\chi^{2}}{TSS}, \nonumber \\
TSS = \sum_{i=1}^{N}w_{i}\left(y_{i}-\overline{y}\right)^{2},\ \overline{y}=\frac{\sum_{i=1}^{N}w_{i}y_{i}}{\sum_{i=1}^{N}w_{i}}.
\end{eqnarray} In above relations, $f$ is the predicted value from the fit, $\beta_{j}$ are fit parameters ($j=1,...,p$) 
and $\overline{y}$ is the weighted mean of the observed data $y_{i}$ at particular $x_{i}$, $i=1,...,N$. $w_{i}$ are the 
weights applied to each data point. $TSS$ is the so called total sum of squares. It is worth stressing that in the case 
of non-linear models, the number of degrees of freedom is generally unknown, i.e.\@ it is not possible to compute the 
value of reduced $\chi^{2}$ or adjusted $R^{2}$ (Andrae et al.\@ 2010). In Table 1, parameter $k =  N - p$ is given for 
convenience. In non-linear models, $k$ does not always represent the exact number of degrees of freedom (Andrae et al.\@ 2010).

\begin{figure}
\plotone{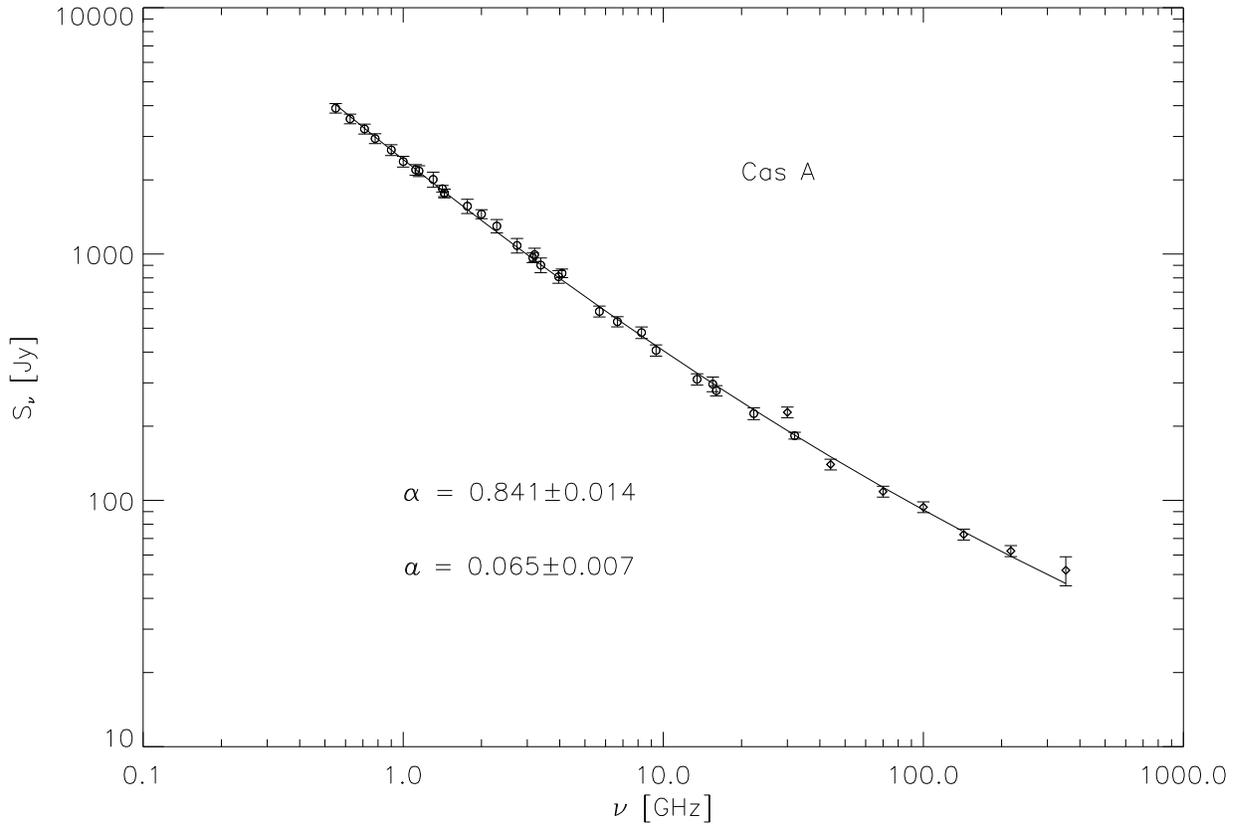}
\caption{The weighted least-square fit to the data from the first sample, for the varying spectral index model. 
Scaling relation taken from Vinyaik\u{\i}n (2014) was used. Diamond symbols indicate {\it Planck's} data}\end{figure}

There are no significant differences in the fit results for both samples and for different scaling relations. 
Again, the exclusion of the data point at 353 GHz does not change the results significantly. It must be noted that 
the smoothly curved spectral shape represented by equation (4) gives a better fit than the model that assumes two 
power laws.

In Figure 2, the weighted least-square fit to the data from the first sample, for the varying spectral index model is 
presented. Scaling relation taken from Vinyaik\u{\i}n (2014) was used. Diamond symbols indicate {\it Planck's} data.

It is also appropriate to check if obtained values of above mentioned parameters, $\alpha$ and $a$, are consistent 
with the ones obtained when low-frequency cut-off is taken into account. To this end, the previous model was adjusted 
in the following way \begin{equation}
S_{[{\rm Jy}]}(\nu) = S_{[{\rm Jy}]}(1\ \!\!\mathrm{GHz})\ \nu_{[{\rm GHz}]}^{-\alpha + a\log\nu_{[{\rm GHz}]}}\ e^{-\tau_{0} \nu_{[{\rm GHz}]}^{-2.1}},                     
\end{equation} and is of the same form as the one stated in Vinyaik\u{\i}n (2014). $\tau_{0}$ represents the optical 
depth at 1 GHz.

For this analysis new data samples were formed and analyzed. The third sample includes data from Tables 3 and 4 from 
Vinyaik\u{\i}n (2014) for the epoch 2015.5 as well as the {\it Planck's} data from Arnaud et al.\@ (2014) appropriately 
scaled to match the same epoch. The frequency range for this sample is 0.0056 - 353 GHz. The fourth sample includes 
all the data from Table 2 presented in Baars et al.\@ (1977) with addition of data from Arnaud et al.\@ (2014), 
Helmboldt \& Kassim (2009) and Mason et al.\@ (1999). Finally, the fifth sample incorporates the fourth one with 
the addition of data from Hurley-Walker et al.\@ (2009) and Liszt \& Lucas (1999). Data in fourth and fifth sample 
are scaled to the epoch of {\it Planck's} observations and the frequency range for these samples is 0.01005 - 353 GHz. 
In all these cases, scaling relation from Vinyaik\u{\i}n (2014) was only used.

\begin{figure}
\plotone{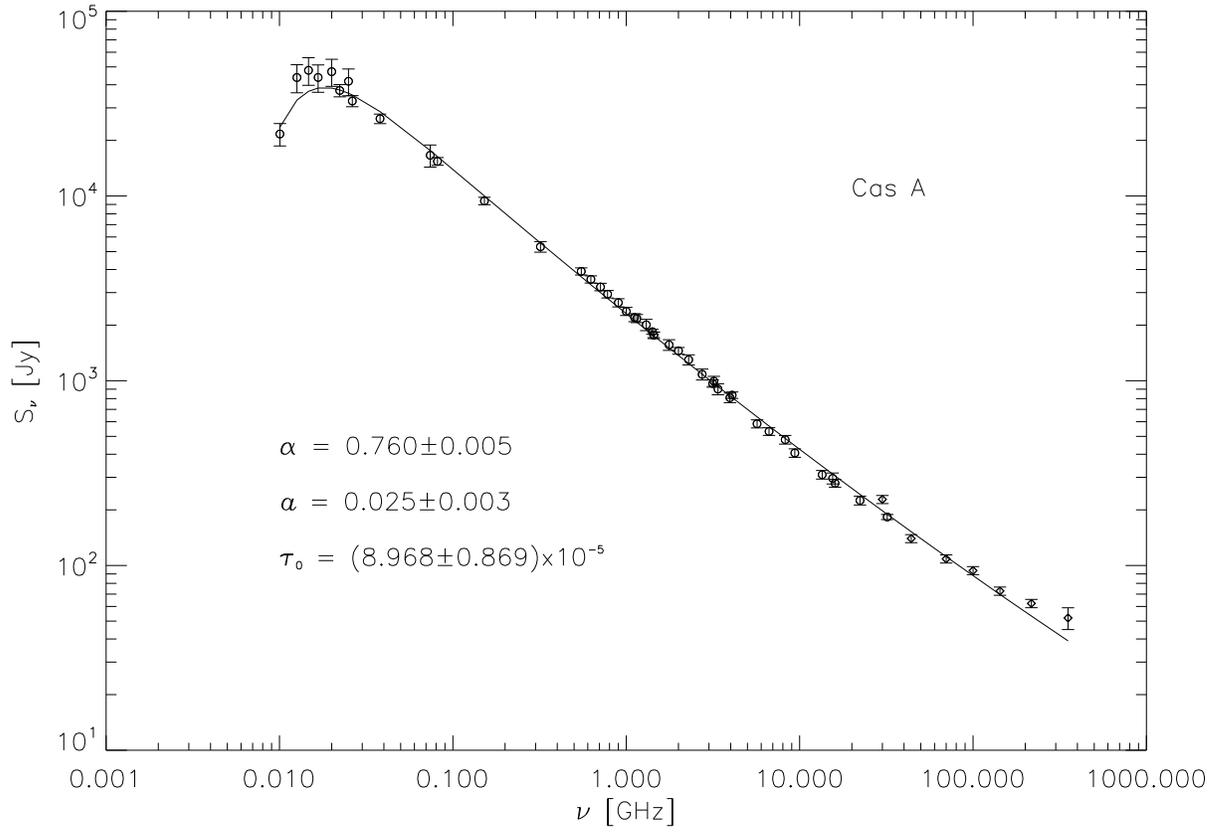}
\caption{The weighted least-square fit to the data from the fourth sample, for the varying spectral index model 
with low-frequency cut-off. Diamond symbols indicate {\it Planck's} data}\end{figure}

\begin{deluxetable}{@{}ccccccc}
\tabletypesize{\scriptsize}
\tablecaption{Varying power-law with low-frequency cut-off best fitting parameters.}
\tablewidth{0pt}
\tablehead{\colhead{sample} & \colhead{frequency range\ [GHz]} & \colhead{$\alpha$} & \colhead{$a$} & \colhead{$\tau_{0}$} & \colhead{$\chi^{2}\ (k)$} & \colhead{$R^{2}$}}
\startdata
3 &$0.0056 - 353$&$0.762\pm0.005$&$0.033\pm0.003$&$(7.738\pm0.574)\times10^{-5}$&$83.05\ (29)$&$0.994$\\
4 &$0.01005 - 353$&$0.760\pm0.005$&$0.025\pm0.003$&$ (8.968\pm0.869)\times10^{-5}$&$75.79\ (46)$&$0.994$\\
5 &$0.01005 - 353$&$0.765\pm0.005$&$0.026\pm0.003$&$(9.209\pm0.875)\times10^{-5} $&$111.03\ (54)$&$0.991$\\
\enddata
\end{deluxetable}

The results of weighted least-square fits to data in this case are summarized in Table 2. There are no 
significant differences in the fit results for all samples. In Figure 3 the weighted least-square fit to the 
data from the fourth sample, for the varying spectral index model with low-frequency cut-off is presented. 
Diamond symbols indicate {\it Planck's} data.

The results are not significantly different than estimated in Vinyaik\u{\i}n (2014). On the other hand, 
the estimated values for spectral index around 1 GHz as well as for the curvature parameter are less than the 
ones obtained using the relation (4) on high frequency part of the radio-spectra alone.

The apparent excess radiation at 353 GHz, discussed in Arnaud et al.\@ (2014), is also present in this analysis 
(see Figure 3). It may be due to the presence of cool dust in northern and western parts of Cas A 
(Dunne et al.\@ 2009; Vinyaik\u{\i}n 2014). It should be noted that images at 600 GHz observed with the Herschel 
Space Observatory (Barlow et al.\@ 2010) at much higher angular resolution than {\it Planck} ($6''$ - $37''$) show 
that the non-synchrotron microwave emission may be a combination of both cold interstellar dust and freshly formed dust. 
Arnaud et al.\@ (2014) stressed that this excess could potentially be due to a coincidental peak in the unrelated 
foreground emission or to cool dust in the supernova remnant, which is marginally resolved by {\it Planck}.

The spinning-dust emission (electric dipole radiation from the rapidly rotating dust grains) could generally contribute 
significantly at high radio-frequencies, especially around 10 - 100 GHz (Draine \& Lazarian 1998ab; 
Ali-Ha\"{\i}moud et al.\@ 2009; Scaife et al.\@ 2007; Silsbee et al.\@ 2011; Stevenson 2014). On the other hand, 
in the case of Cas A this emission component is negligible.

Finally, Vinyaik\u{\i}n (2014) showed that the observed slowing of the secular variations of the radio flux density of Cas A 
with decreasing frequency at decimeter wavelengths can be explained by a decrease in the optical depth of a remnant 
\mbox{H\,{\sc ii}} zone around Cas A with time due to recombination of hydrogen atoms. Due to the presence of the 
\mbox{H\,{\sc ii}} region it may be interesting to fit the radio spectrum of Cas A with the following 
expression \begin{eqnarray}
S(\nu) = S_{1}\ \nu^{-\alpha}\ e^{-\tau_{\nu}} + S_{2}\ \nu^{2}\left(1-e^{-\tau_{\nu}}\right), \nonumber \\
\tau_{\nu} = \tau_{0} \nu^{-2.1}.                    
\end{eqnarray} This formula incorporates both synchrotron radiation as well as the thermal absorption 
and thermal bremsstrahlung emission from the zone in front of the synchrotron emitting region. Of course, this 
is a rather naive model. Supernova remnants are 3D structures, the fact that this relation does not take into 
account. On the other hand, it can serve as an approximative model for qualitative analysis.

If frequencies are in GHz then the (non-thermal) synchrotron flux density at 1 GHz is simply 
$S^{\rm NT}(1\ \!\!\mathrm{GHz}) = S_{1}$ but for the thermal component holds 
\begin{eqnarray}
S^{\rm T}(1\ \!\!\mathrm{GHz}) = S_{2}\left(1-e^{-\tau_{0}}\right), \nonumber \\
\Delta S^{\rm T}(1\ \!\!\mathrm{GHz}) = S^{\rm T}(1\ \!\!\mathrm{GHz}) \left(\frac{\Delta S_{2}}{S_{2}} + \Delta\tau_{0}\ \frac{1}{e^{\tau_{0}}-1}\right),
\end{eqnarray}
so that the contribution of thermal emission in integral radiation at 1 GHz is given by
\begin{eqnarray}
\xi=\frac{S^{\rm T}(1\ \!\!\mathrm{GHz})}{S^{\rm T}(1\ \!\!\mathrm{GHz}) + S^{\rm NT}(1\ \!\!\mathrm{GHz})}, \nonumber \\
\Delta\xi=\xi\left(\frac{\frac{\Delta S^{\rm T}(1\ \!\!\mathrm{GHz})}{S^{\rm T}(1\ \!\!\mathrm{GHz})} + \frac{\Delta S^{\rm NT}(1\ \!\!\mathrm{GHz})}{S^{\rm NT}(1\ \!\!\mathrm{GHz})}}{1+\frac{S^{\rm NT}(1\ \!\!\mathrm{GHz})}{S^{\rm T}(1\ \!\!\mathrm{GHz})}}\right).
\end{eqnarray} The results of weighted least-square fits to data in this case are summarized in Table 3. Again, there 
are no significant differences in the fit results for all samples. The weighted least-square fit to the data from the fourth 
sample, for the test-particle synchrotron model with thermal absorption and emission is presented in Figure 4. 
Diamond symbols indicate {\it Planck's} data.

\begin{figure}
\plotone{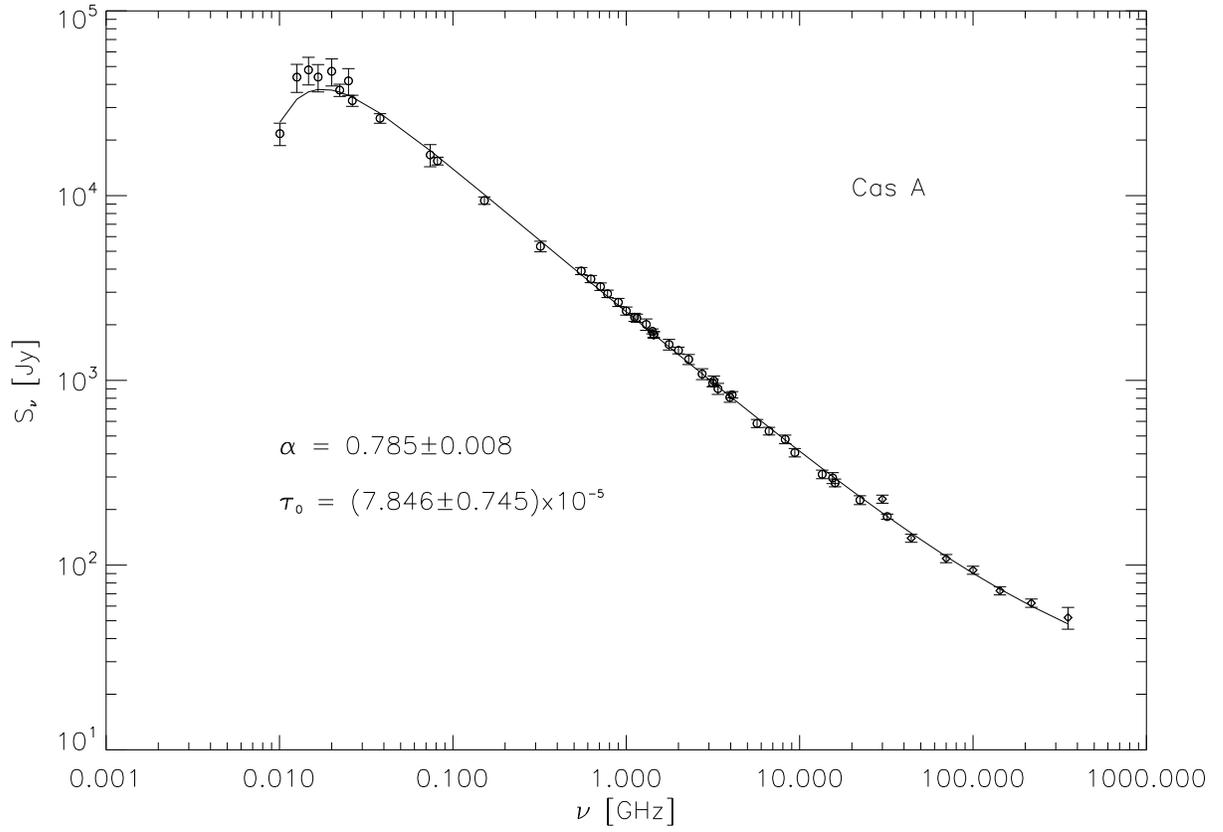}
\caption{The weighted least-square fit to the data from the fourth sample, for the test-particle synchrotron model 
with thermal absorption and emission. Diamond symbols indicate {\it Planck's} data}\end{figure}

\begin{deluxetable}{ccccccc}
\tabletypesize{\scriptsize}
\tablecaption{Test-particle synchrotron model with thermal absorption and emission best fitting parameters.}
\tablewidth{0pt}
\tablehead{\colhead{sample} & \colhead{frequency range\ [GHz]} & \colhead{$\alpha$} & \colhead{$\tau_{0}$} & 
\colhead{$\xi\ [\%]$} & \colhead{$\chi^{2}\ (k)$} & \colhead{$R^{2}$}}
\startdata
3 &$0.0056 - 353$&$0.769\pm0.006$&$(5.211\pm0.341)\times10^{-5}$&$1.653\pm0.005$&$53.97\ (29)$&$0.996$\\
4 &$0.01005 - 353$&$0.785\pm0.008$&$(7.846\pm0.745)\times10^{-5}$&$1.913\pm0.008$&$47.35\ (46)$&$0.996$\\
5 &$0.01005 - 353$&$0.788\pm0.008$&$(7.535\pm0.624)\times10^{-5}$&$1.887\pm0.007$&$74.12\ (54)$&$0.994$\\
\enddata
\end{deluxetable}
The presence of thermal emission from the \mbox{H\,{\sc ii}} region associated with this SNR is estimated to be around 
1-2 \% in the total flux density at 1 GHz. Addition of this component significantly improves fit to the data but discards 
the curvature due to non-linear effects of particle acceleration. It is worth noting that addition of the curvature 
parameter $a$ to the previous model, i.e.\@ varying power-law model with thermal absorption and emission in front 
of synchrotron emitting region, does not produce acceptable fits to all samples. Negative, i.e.\@ non-physical in 
sense of the non-linear DSA effects\footnote{When the pressure of cosmic-ray (CR) particles, produced at the shock wave, 
is included in the analysis, such a modification of shock structure implies changes in the spectrum of accelerated electrons. 
The energy spectrum of low energy electrons becomes softer (the radio spectrum is steeper) and the spectrum of high energy 
electrons becomes harder (the radio spectrum is shallower) than that of linear DSA. Due to significant CR production, the 
concave-up (positively curved) radio spectra is expected in the case of young SNRs (see Uro\v sevi\'c 2014, and references 
therein for more details).}, values for the curvature parameter $a$ are preferred. Bounding the parameters to appropriate 
physical ranges of values do not improve fits that are also highly sensitive to initial guesses for the parameters. On the 
other hand, simple model represented by equation (7) fits to data very well (see Figure 4). The flux density estimate at 
353 GHz is within the error bars of the measured value at this frequency for data samples 4 and 5 (see Figure 4).

\section{Discussion}

It is known that some young SNRs, such as Cas A and Kepler exhibit evidence of radio spectral variations from one region 
to another (Anderson \& Rudnick 1996; DeLaney et al.\@ 2002). This can lead to a curved composite spectrum even if the 
synchrotron spectrum for each region is just a power-law (Allen et al.\@ 2008). On the other hand, the results of 
Jones et al.\@ (2003) indicate that the spectra of particular small features in Cas A flatten with increasing energy 
and Allen et al.\@ (2008) found that their results can be explained with model that includes curvature of particle spectra 
of $b=0.06\pm0.01$. Curvature of particle spectra $b$ is in fact four times spectral curvature $a$ so that our 
results, although slightly higher, are not in disagreement with conclusions of Allen et al.\@ (2008).

For additional support to our claims, we used the same model for synchrotron radiation as in Allen et al.\@ (2008), 
with one simplification that the non-thermal particle distribution function lacks high energy cut-off \begin{equation}
N(E)dE=K\left(\frac{E}{E_{0}}\right)^{-\Gamma+b\log\frac{E}{E_{0}}},\end{equation}
where $b=4a$ is the curvature of particle spectra, $E_{0}=1\ {\rm GeV}$ and $\Gamma$ is particle spectral index at $E_{0}$. 
Flux density is a combination of synchrotron radiation from such an ensemble and thermal absorption at low frequencies in 
a similar manner as in equation (6). The results, given in Table 4, show that the simplified model based on the one from 
Allen et al.\@ (2008) is consistent with our previous results. 

\begin{deluxetable}{@{}ccccccc}
\tabletypesize{\scriptsize}
\tablecaption{Simplified model based on the one from Allen et al.\@ (2008) best fitting parameters.}
\tablewidth{0pt}
\tablehead{\colhead{sample} & \colhead{frequency range\ [GHz]} & \colhead{$\alpha$} & \colhead{$a$} & \colhead{$\tau_{0}$} 
& \colhead{$\chi^{2}\ (k)$} & \colhead{$R^{2}$}}
\startdata
3 &$0.0056 - 353$&$0.775\pm0.007$&$0.032\pm0.001$&$(8.032\pm0.500)\times10^{-5}$&$79.57\ (28)$&$0.994$\\
4 &$0.01005 - 353$&$0.739\pm0.005$&$0.025\pm0.003$&$ (8.998\pm0.867)\times10^{-5}$&$75.72\ (45)$&$0.994$\\
5 &$0.01005 - 353$&$0.777\pm0.008$&$0.026\pm0.003$&$(9.268\pm0.872)\times10^{-5} $&$110.90\ (53)$&$0.991$\\
\enddata
\end{deluxetable}

Finally, Dunne et al.\@ (2009) showed that polarized sub-millimeter emission is associated with the SNR and 
that the excess polarized sub-millimeter flux at 353 GHz is due to cold dust within the remnant. They noted 
that there is no currently known way to produce such a polarized sub-millimeter emission from a synchrotron 
process.

Keeping in mind the fact that inclusion of thermal bremsstrahlung emission (equation 7) apparently leads to 
a better fit to data (Table 3), it must be noted that such an interpretation leads to a less contribution 
of dust emission at 353 GHz, which is in contrast to the conclusion of Dunne et al.\@ (2009). In that sense, 
we conclude that the more probable scenario is that {\it Planck's} new data actually reveal the imprint of 
non-linear particle acceleration in the case of this young Galactic SNR. Of course, we can not fully 
dismiss the possible contamination by adjacent \mbox{H\,{\sc ii}} region.

\begin{figure}
\plotone{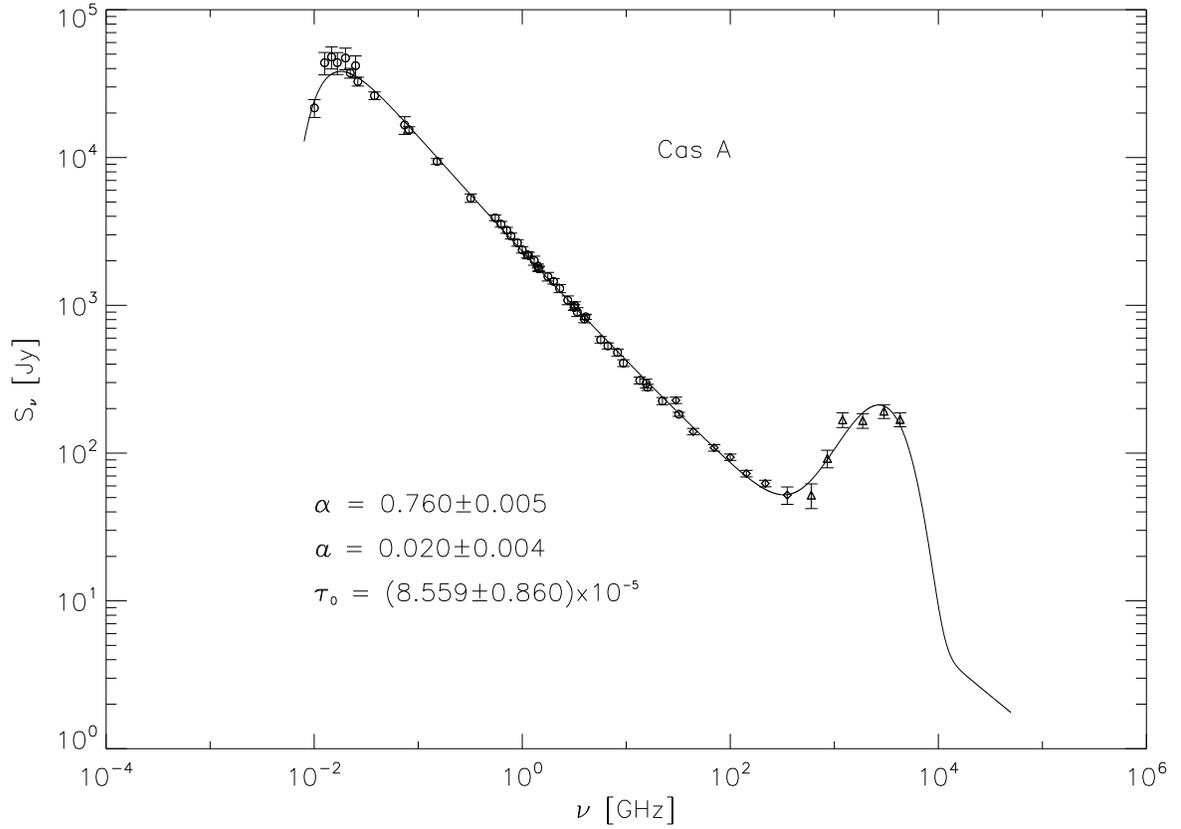}
\caption{The weighted least-square fit to the data from the seventh sample, for the non-linear synchrotron model 
with thermal absorption and dust emission. Diamond symbols indicate {\it Planck's} and triangles {\it Herschel's} 
data}\end{figure}

The additional attempt to support the non-linear particle acceleration hypothesis may come from the fit using 
non-linear synchrotron model with thermal absorption and dust Planck-like emission (Figure 5 and Table 5). 
To that end, the {\it Herschel's} total flux densities from Barlow et al.\@ (2010) were added to data samples 
3-5 forming samples 6-8. Again, similar results are obtained. Of course, this is a simplified model which does 
not distinguish between different dust emission components. On the other hand, the spectral curvature due to 
the non-linear particle acceleration is clearly present in this case too (see Table 5). These values for 
$a$ are also more in accordance with the result of Allen et al.\@ (2008).

\begin{deluxetable}{@{}ccccccc}
\tabletypesize{\scriptsize}
\tablecaption{The non-linear synchrotron model with thermal absorption and dust emission best fitting parameters.}
\tablewidth{0pt}
\tablehead{\colhead{sample} & \colhead{frequency range\ [GHz]} & \colhead{$\alpha$} & \colhead{$a$} & \colhead{$\tau_{0}$} 
& \colhead{$\chi^{2}\ (k)$} & \colhead{$R^{2}$}}
\startdata
6 &$0.0056 - 4285.714$&$0.756\pm0.005$&$0.022\pm0.004$&$(6.807\pm0.579)\times10^{-5}$&$60.86\ (33)$&$0.995$\\
7 &$0.01005 - 4285.714$&$0.760\pm0.005$&$0.020\pm0.004$&$ (8.559\pm0.860)\times10^{-5}$&$69.55\ (50)$&$0.994$\\
8 &$0.01005 - 4285.714$&$0.764\pm0.005$&$0.021\pm0.003$&$(8.800\pm0.865)\times10^{-5} $&$102.01\ (58)$&$0.992$\\
\enddata
\end{deluxetable}

It is also worth mentioning that Pohl et al.\@ (2015) showed that stochastic re-acceleration of 
electrons downstream of the forward shock can explain the soft spectra observed from many Galactic SNRs. 
They also noted that, generally, interiors of the SNRs produce slightly steeper radio spectra than does the 
shell where re-acceleration occurs.  

\section{Conclusions}

The main conclusions that can be drown from this analysis are

(1) {\it Planck's} data support the observation that the radio continuum of SNR Cas A flattens with increasing 
frequency. The spectrum becomes positively curved above around 30 GHz.

(2) Alternative explanations of curvature are investigated: significant non-linear effects of particle acceleration, 
presence of thermal emission due to the emission of associated \mbox{H\,{\sc ii}} region, and the presence of 
dust emission.

(3) Different samples of data were analyzed to account for differences in data gathered from various literature. The 
presented results are not significantly sensitive to differences between samples.

(4) The results presented in this paper agree with the conclusions stated in Vinyaik\u{\i}n (2014), i.e.\@ the observed 
flattening of the radio spectrum at millimeter wavelengths is most likely primarily due to a flattening of the radio 
synchrotron spectrum of Cas A itself. In other words, non-linear effects of particle acceleration are possibly mainly 
responsible for the apparent high-frequency curvature in Cas A radio spectrum.

\acknowledgments

We wish to thank the anonymous referee for useful suggestions which substantially improved this paper.
This work is part of Project No. 176005 "Emission nebulae: structure and evolution" supported by the Ministry 
of Education, Science, and Technological Development of the Republic of Serbia.

\end{document}